\documentclass[lettersize,journal]{IEEEtran} 
\usepackage{amsmath,amsfonts,amssymb}
\usepackage{algorithmic}
\usepackage{algorithm}
\usepackage{array}
\usepackage[caption=false,font=normalsize,labelfont=sf,textfont=sf]{subfig}
\usepackage{textcomp}
\usepackage{stfloats}
\usepackage{url}
\usepackage{verbatim}
\usepackage{graphicx}
\usepackage{cite}
\usepackage{hyperref}
\usepackage{orcidlink}
\usepackage{siunitx}
\usepackage{xcolor}

\begin{document}

\title{On the Development of an RFSoC-Based Ultra-Fast Phasemeter With GHz Bandwidth}
\author{Shreevathsa Chalathadka Subrahmanya\orcidlink{0000-0002-9207-4669}, Christian Darsow-Fromm\orcidlink{0000-0001-9602-0388}, and Oliver Gerberding\orcidlink{0000-0001-7740-2698}
\thanks{
This work was supported in part by the Deutsche Forschungsgemeinschaft (DFG, German Research Foundation) under Germany’s Excellence Strategy—EXC 2121 “Quantum Universe” under Grant 390833306, in part by the Deutsches Zentrum f\"ur Luft- und Raumfahrt (DLR) with funding from the Bundesministerium f\"ur Wirtschaft und Klimaschutz under Project 50OQ2001 and Project 50OQ2302, and in part by the Open Access Publication Fund of Universit\"at Hamburg. \textit{(Corresponding author: Shreevathsa Chalathadka Subrahmanya.)}}%
\thanks{The authors are with the Institute of Experimental Physics, University of Hamburg, 22761 Hamburg, Germany. (e-mail: \href{mailto:shreevathsa.subrahmanya@uni-hamburg.de}{shreevathsa.subrahmanya@uni-hamburg.de}; \href{mailto:oliver.gerberding@uni-hamburg.de}{oliver.gerberding@uni-hamburg.de}).}}


\maketitle

\begin{abstract}
Precise measurements of the change in the frequency and phase of an electrical or optical signal play a key role in various branches of science and engineering.
Tracking changing laser frequencies is especially demanding when the lasers themselves are noisy or if the frequencies rapidly change because they encode highly dynamic signals in, e.g., Doppler-ranging or dynamic cavity readout.
Here, to address these and other possible applications, we report the development of a high signal bandwidth (\textgreater\,2\,GHz) and high tracking bandwidth (2\,MHz) multi-channel phasemeter.
The implementation utilizes an all-digital phase-locked loop realized within the field programmable gate array (FPGA) part of a radio frequency system-on-chip (RFSoC), the programmable logic (PL).
The benchmark level parameters of the phasemeter are obtained by operating the PL at its timing limits and introducing a ovel multi-demodulation and phase accumulation scheme.
We present performance measurements, discuss the role of the high tracking bandwidth for tracking highly dynamic signals, and demonstrate ultra-stable phase locking of a beat note between two widely tunable external cavity diode lasers.
Within our implementation, we have included a direct measurement of the open-loop gain of the phase-locked loop (PLL) and an estimation of the residual phase error.
These features are new in the field of FPGA-based phasemeter developments.
We achieve a phase-noise floor in the sub-milli radian regime when comparing two signals, even for GHz frequencies, and demonstrate stable tracking of signals with a frequency change rate of 240\,GHz/s.
\end{abstract}

\begin{IEEEkeywords}
    Field programmable gate arrays (FPGAs), frequency measurement, phase-locked loops (PLLs).
\end{IEEEkeywords}

\section*{Nomenclature}
\addcontentsline{toc}{section}{Nomenclature}
\begin{IEEEdescription}[\IEEEusemathlabelsep\IEEEsetlabelwidth{ADPLL}]
\item[ADC] Analog\textendash digital converter.
\item[ADPLL] All-digital phase-locked loop.
\item[CIC] Cascaded integrator-comb.
\item[FPGA] Field programmable gate array.
\item[LISA] Laser interferometer space antenna. 
\item[NCO] Numerically controlled oscillator.
\item[PIR] Phase increment register.
\item[PL] Programmable logic.
\item[PLL] Phase-locked loop.
\item[RFSoC] Radio frequency system-on-chip.
\item[rms] Root-mean-square.
\end{IEEEdescription}

\section{Introduction}
\IEEEPARstart{M}{easurement} and tracking the changing frequency or phase of a radio frequency (RF) electrical signal are essential in many precision experiments and applications.
The RF signals encode various side-bands that carry information about the motion of a target, changes in a cavity mode, the stability of the carrier generating oscillator, or information intentionally modulated onto the carrier in, e.g., telecommunications. 

We classify the instrument that measures and tracks the phase and frequency of an incoming electrical signal as a phasemeter. There are different methods to realize such measurements, such as zero-crossing~\cite{Pollack2006} and frequency counting~\cite{6230960}, but here we focus on phase-tracking with a digital local oscillator, which provides the highest possible signal-to-noise ratio.
A PLL locks a local oscillator to an incoming signal, providing the necessary phase information.  
To measure phase and frequency with high precision, digitizing the signal and embedding PLLs in FPGAs is a well-studied approach~\cite{Shaddock2006,Gerberding2013}. The core algorithm of such a digital phasemeter is an ADPLL. 

Such digital phasemeters have been developed, e.g., for laser interferometry, prominently in the context of the space-based gravitational-wave observatory, the LISA~\cite{Shaddock2006}. Phasemeters are also available as commercial instruments~\cite{moku}. 

The developed phasemeter is meant to track one or multiple tones in the input signal and thereby give information on how the phase and/or frequency of that tone changes over time.
Determining the absolute phase of the measured tones relative to some reference is not the main observable of the current instrument. Such absolute phase information can be measured with our device but requires a calibration of the phase/frequency response (or transfer function), which we have not yet studied for our system but has been accomplished in the literature~\cite{Tudosa2022,Chaudhary2015}. Effects due to a non-calibrated phase response are important as second-order disturbances for laser interferometers, prominently for the readout of beam tilts in differential wavefront sensing, and correcting the phase variations measured from tones with quickly changing frequencies.

PLL phasemeters, as discussed here, are distinguished from general signal processing instruments like oscilloscopes or spectrum analyzers. While the latter instruments do provide frequency measurements of (stable) tones, they are not able to measure quickly changing frequencies, their measurements are not available with low-latency and high sampling rate, and they do not achieve precision levels in the milli to micro-Hertz/radian regime that is required for laser interferometry.

In LISA, the RF signal is generated  by beating two lasers with an effective carrier wavelength of about \SI{1}{\micro\meter}, with one beam carrying the Doppler shift associated with the distance changes between two spacecraft.
The beat note frequencies are in the range of \qtyrange[range-units=single,range-phrase=--]{5}{25}{\mega\hertz}, corresponding to speeds of \qtyrange[range-units=single,range-phrase=--]{5}{25}{\meter/\s}, with a maximum rate of frequency change in the \SI{20}{\hertz/s} range~\cite{Schwarze2019}. A phase change of \SI{1}{\micro\radian} for the laser interferometer corresponds to a length change of about \SI{170}{\femto\meter}, indicating the high displacement measurement precision achievable when deploying a digital phasemeter with \textmu rad readout noise floor.

The tracking (or loop) bandwidth of digital phasemeters is one of their core design parameters. It is determined by the open-loop gain of the PLL and hence defines the frequency range up to which we get closed-loop suppression from the PLL. Tracking bandwidth determines whether the phase lock is stable and the loop operates linearly or if non-linear effects are present, resulting in non-linear phase and amplitude measurements, cycle slips, or loss of lock~\cite{1095423}.

High tracking bandwidth is desired for highly dynamic signals that are minimally spoiled by additive noise. Since laser interferometry uses short wavelengths, there are use cases where the dynamics of the frequency changes would be high, motivating the study of phasemeters with as high tracking bandwidth as possible~\cite{Hsu2010}. Also, with higher tracking bandwidth, one can use lasers with higher frequency noise and line width, which enables the use of less stable laser types and lasers with lower costs, size, and weight. 

The other extreme, low tracking bandwidth, is helpful when tracking RF signals spoiled by additive noise, like shot noise for laser interferometers with very low received power. In this case, the signal dynamics have to be minimized by, e.g., stabilizing the used laser systems~\cite{Sambridge2023}; otherwise, the ADPLL will not be able to track the phase dynamics sufficiently. In this paper, we focus only on the high tracking bandwidth case.

The second critical parameter of digital phasemeters is their signal (or detection) bandwidth, the range of RF frequencies that can be tracked. This, in turn, limits the maximum relative speeds that can be tracked in applications such as Doppler-ranging. The ability of digital phasemeters to measure frequencies with more than ten orders of magnitude of dynamic range, as demonstrated for LISA~\cite{Schwarze2019}, can be utilized to enable ultra-linear measurements in a variety of frequency ranges. The lower limit of this dynamic frequency measurement range is the phase readout noise floor, and the upper limit is the detection bandwidth of the phasemeter. With higher detection bandwidth implementations, more applications can use the exquisite linearity of digital phasemeters.

A specific use case we consider for our study is tracking a laser frequency, where the laser is locked to an optical cavity and interfered with another stable reference laser. If one of the mirrors of the cavity is a movable target, then the frequency encodes the displacement of this target~\cite{vanHeijningen2023}.  
For example, a \SI{5}{cm} long optical cavity with a free-spectral range of \SI{3}{\giga\hertz} will, if probed with a \SI{1}{\micro\meter} wavelength laser, lead to \SI{6}{\giga\hertz/\micro\meter} sensitivity. 
That means tracking the frequency change with \unit{\milli\hertz} precision corresponds to a displacement readout noise contribution of the phasemeter of $< 10^{-18}$\,\unit{\meter}. (In practical implementations, other noise sources will limit this to higher levels~\cite{Eichholz2015}.) 
The tracking and detection bandwidths of the digital phasemeter used to track the frequency will determine the maximum tolerable range and speed of mirror motion in the cavity.
This type of displacement sensing is particularly desired in the context of ground and space-based gravitational-wave detectors, with the readout of inertial sensors as the prime application~\cite{vanHeijningen2023,Carter2020,Hines2023}.

In this paper, we present the implementation and testing of a digital phasemeter with benchmark levels of tracking and detection bandwidths. 
To this end, the phasemeter is realized by adapting and optimizing the ADPLL architecture in PL at the hardware limits.
It is worth noting that we are using the phasemeter technologies developed in the context of space-based laser interferometric experiments as a starting point.
We see the potential of a high-bandwidth ultra-fast phasemeter in the fields mentioned before and have developed such a phase measurement instrument.

The maximum achievable tracking bandwidth of phasemeters has not been discussed in detail previously. This is because a lower bandwidth was optimal for LISA~\cite{Gerberding2013}, but it is inherently limited by the clock speed of the PL. As for the signal bandwidth, several previous studies cover different ranges, and we give a short overview of those developments here.
The LISA phasemeter is designed to have a signal bandwidth above \SI{25}{\mega\hertz}, with most implementations providing \SI{40}{\mega\hertz}~\cite{Shaddock2006}, and similar values are required for other space-based missions such as Taiji~\cite{Liu2021}.
In \cite{Liu2014}, the signal bandwidth was limited to \qtyrange[range-units=single,range-phrase=--]{0.5}{10}{\mega\hertz} range because of the considered design.
For a high-speed multiplexed heterodyne interferometry, a phase measurement system with a signal bandwidth of \SI{156.25}{\mega\hertz} was demonstrated~\cite{Isleif2014}.
In a heterodyne laser frequency stabilization experiment, a LISA phasemeter-like system was used but with a modified signal bandwidth of \SI{32}{\mega\hertz}~\cite{Eichholz2015}.
The maximum bandwidth offered by the commercially available phasemeter is \SI{300}{\mega\hertz}~\cite{moku}.
Compared to all these available and previous developments, our phasemeter provides a \SI{2.048}{\giga\hertz} signal bandwidth while maintaining the measurement noise floor in the range of the LISA requirement~\cite{Shaddock2006}.

In Sec.~\ref{sec:implementation}, we describe the specific implementation of our ADPLL within the hardware limitations, and we present and verify a linear loop and noise model. 
This also includes the addition of a new readout signal from the ADPLL, which gives us a real-time probe of its linearity that we can use to optimize the loop gains. 
We then present performance measurements of our phasemeter in Sec.~\ref{sec:performance}, where we specifically demonstrate the high tracking bandwidth by locking it to the beat note between two noisy lasers. 
Finally, we characterize the acquisition range and tracking speed of the ADPLL. 

\section{Implementation}
\label{sec:implementation} \noindent %
The Zynq UltraScale+ RFSoC ZCU111 evaluation kit~\cite{rfsoc} is the basis for our implementation. 
This RFSoC features eight 12-bit \SI{4.096}{GSPS} analog\textendash digital converters (ADCs), allowing us to implement a maximum of 8 readout channels, each with a signal bandwidth in the GHz regime. 
The PL is controlled, and the data is acquired in real-time via Yocto Linux running on the on-chip Arm Cortex-A53 and Arm Cortex-R5 subsystems.
The RFSoC is programmed directly using the very high speed integrated circuit hardware description language (VHDL).

Our phasemeter provides two ADPLLs per channel, one to track the main tone and the other to track a pilot tone, allowing us to correct for the ADC timing jitter in post-processing~\cite{Shaddock2006}. 
Hence, in its maximum capacity, the ZCU111 evaluation board runs as an eight-channel phasemeter with a total of 16 ADPLLs. 
Each PLL can track a distinct tone and is controlled individually by the processing system by setting registers using a Python interface.

Although the general signal bandwidth of each channel is \SI{2.048}{\giga\hertz}, the XM500 RFMC balun transformer add-on card currently in use limits the first two channels to single-ended signals of frequencies up to \SI{1}{\giga\hertz}, and the next two channels are limited to frequencies above \SI{1}{\giga\hertz}. The remaining four channels are available for the entire signal bandwidth, with differential inputs. 

The IQ-demodulation embedded in a closed-loop phase and frequency tracking system, leading to a general ADPLL, is explained in~\cite{Gerberding2013}. 
However, our implementation of the ADPLL required some adaptations because the highest processing rate of the PL is limited to \SI{512}{\mega\hertz}, which is eight times less than the data acquisition rate of \SI{4.096}{GSPS}. To tackle this limitation, we are introducing the following novel implementation strategy.

The RF data converters are configured to provide 8 data samples in parallel.
To use the full potential of RFSoC, we introduce an extended multi-demodulation scheme followed by a multi-phase accumulation stage, allowing a direct tracking of signals of frequencies up to \SI{2.048}{\giga\hertz} with the processing operating at \SI{512}{\mega\hertz}. Running the PL at such rates is challenging and requires fine-tuning of registers within the loop to achieve the timing requirements that are strained at this speed already from simple routing delays. However, keeping this processing speed is crucial to achieve high tracking bandwidth, which is limited by ADPLL-internal delays, as discussed below. 

\subsection{Working Principle} 
\begin{figure}[!t]
\centering
\includegraphics[width=3.4in]{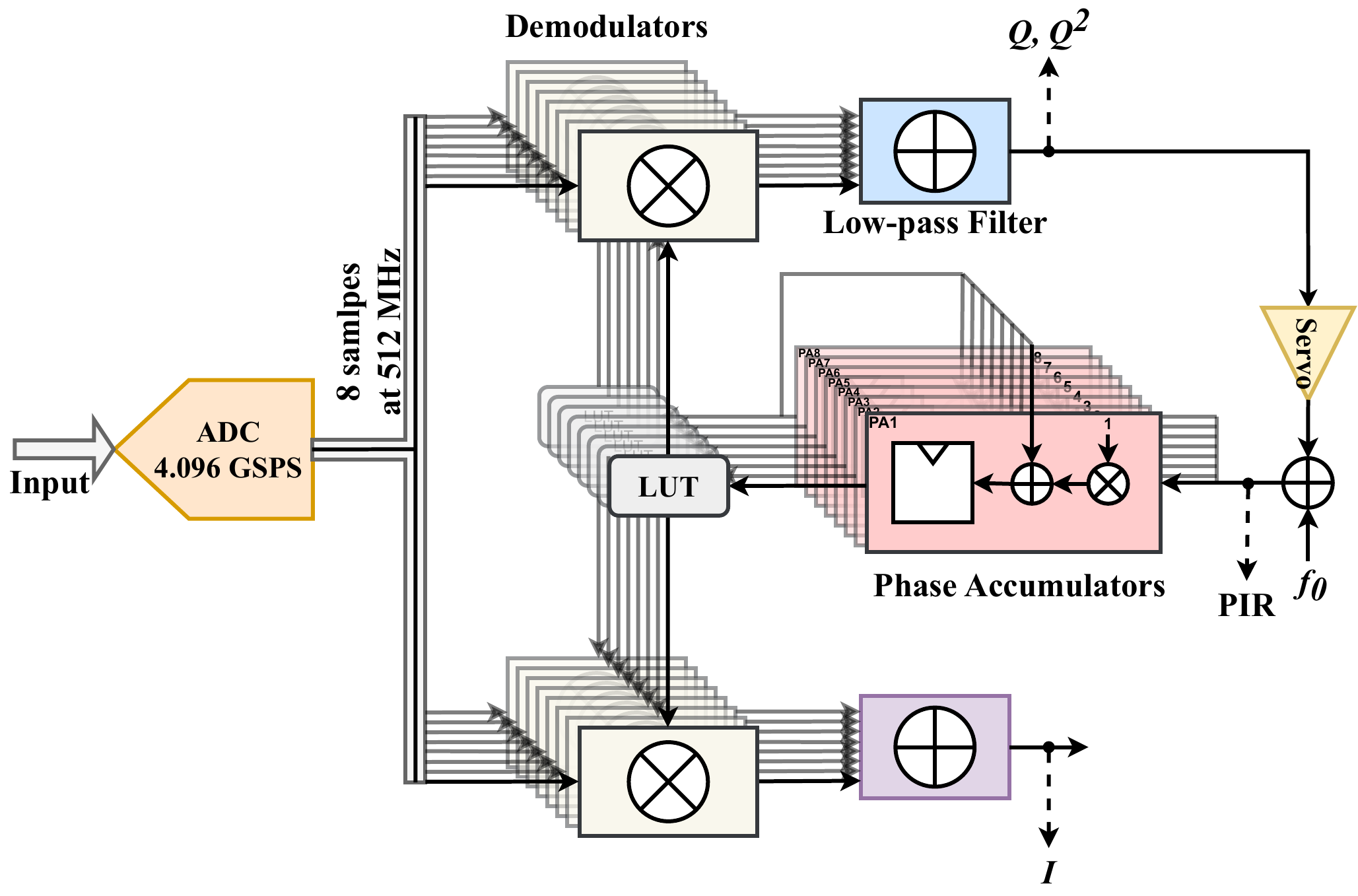}
\caption{The GHz Phasemeter topology, where the incoming electrical signal is digitized at \SI{4.096}{\giga\hertz}, giving out eight 12-bit samples at \SI{512}{\mega\hertz}. 
Markers for signal readout points are shown. PIR: phase increment register, PA: phase accumulator, LUT: look-up table.}
\label{fig1}
\end{figure}

\noindent %
The PLL used in the present work is an all-digital implementation in which a local oscillator, specifically an NCO, tracks the frequency of the incoming signal. 
Tracking is realized by mixing both signals and using the quadrature signal ($Q$) as an error signal in a control loop. 
The overview of the GHz PLL, including multi-demodulation and multi-phase accumulation schemes, is sketched in Fig.~\ref{fig1}.
If one follows the basic topology as in \cite[Fig.~1]{Gerberding2013}, the phasemeter will be limited to a signal bandwidth of \SI{256}{\mega\hertz}. 
However, the proposed multi-demodulation and multi-phase accumulation schemes push the bandwidth to the \unit{\giga\hertz} regime, making it possible to realize such an instrument.
Without the proposed signal processing scheme modifications, achieving reliable operation with \unit{\giga\hertz} regime signal bandwidth using RFSoC would not be possible.

For an ADPLL to get locked to the input signal, a starting frequency $f_0$, which is close to the frequency of the input signal, is specified.  
This frequency value is stored in a phase increment register (PIR). As each ADC channel gives eight data samples for each clock cycle, we need eight consecutive phase values in the NCO for multi-demodulation. Hence, the PIR value is split and then incremented successively, such that an effectively continuous phase value is generated for eight ook-up tables (LUTs).
Then, the LUTs generate sine and cosine signals of the given phase, and along with the previous multi-phase accumulators, they form the NCO of the phasemeter. 
The digitized input signal samples are each then IQ demodulated at NCO frequencies and filtered by a rolling average of 16 samples (the averaging is implemented in consecutive additions to account for timing constraints). The so-generated $Q$ value acts as the error signal and is fed into the servo (Proportional-Integral controller).
Its output is the actuation signal of the control loop that corrects the initial starting frequency to match the input signal frequency. 

When the PLL is locked, reading out the PIR gives us a copy of the frequency of the input electrical signal. $Q$ measures how good the tracking is, and $I$ scales to the amplitude of the input signal. 
The readout of the ADPLLs is realized with second-order CIC filters (except for $Q^2$, as discussed below) that decimate the data to a desired rate between \SI{10}{\kilo\hertz} and \SI{30.5}{\hertz}.
Currently, the upper limit is set by the software in the on-board processing subsystem, which could be increased by optimizing the performance of the readout code.

\begin{figure*}[!t]
\centering
\includegraphics[width=6in]{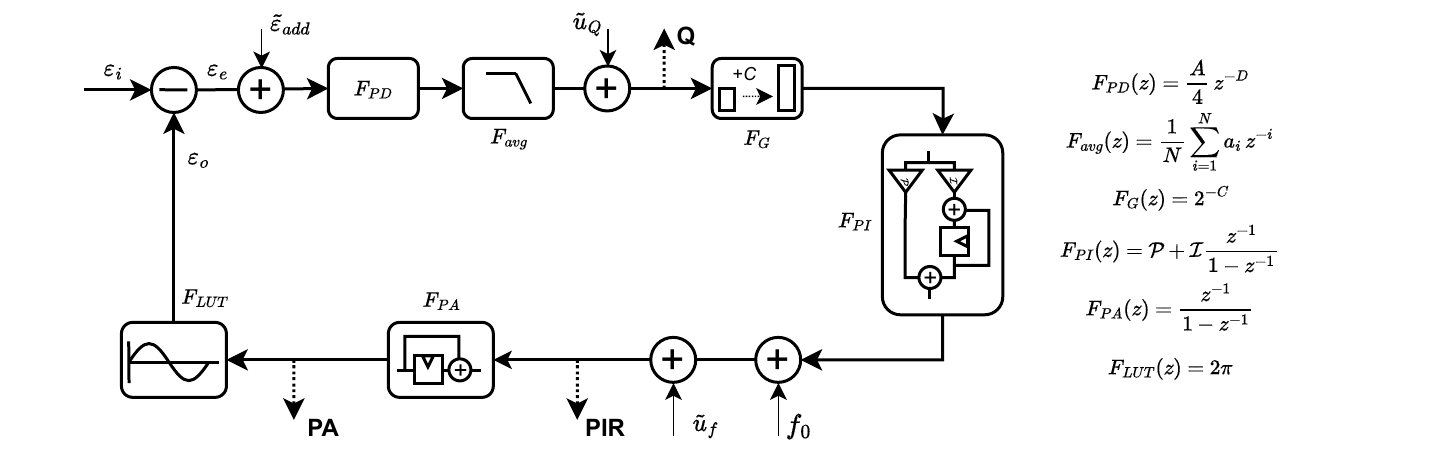}
\caption{Block diagram of the linearized PLL model considering phase ($\varepsilon$) as the sensed and actuated quantity. Input signal amplitude $A$ and a total number of $D$ delays at the sampling frequency, resulting in the computational delay transfer function, are taken into the frequency response of the phase detector ($F_{PD}$). 
Markers for signal readout and possible noise addition points are shown. 
The transfer function of each block in the z-domain is provided.
$\Tilde{\varepsilon}_{add}$: additive phase noise, $\Tilde{u}$: truncation phase noise, $G$: overall gain due to in-loop bit-growth, PI: proportional-integral controller, $f_0$: starting frequency, PA: phase accumulator, LUT: ook-up table.}
\label{fig2}
\end{figure*}

\subsection{ADPLL Model and Loop Bandwidth} \noindent %
When the PLL is locked to the input signal, we expect it to operate in a linear regime that is achieved when $Q$ is small. With this assumption, it is possible to model the PLL reliably by understanding the transfer function of each block in the loop.
A block diagram of our implementation is sketched in Fig.~\ref{fig2}.
For a detailed discussion of the loop model, one can refer to~\cite{Gerberding2013}.
With the transfer function of each block known, the open-loop gain of the modeled PLL is just the multiplication of all these individual stages.

The highest stable loop bandwidth is limited only by the delays within the ADPLL caused by registers, which in turn are necessary to realize the various processing steps. 
The value of the individual delays is determined directly by the processing speed. The number of delays depends on the implementation and its optimization in terms of bit-width. 
For example, the utilization and settings of dedicated digital-signal-processing blocks in the PL allow us to minimize the number of registers in the loop. 
While operating the PL at \SI{512}{\mega\hertz} requires more registers in the loop to meet timing requirements, we still use less than twice the amount of registers needed to operate at \SI{256}{\mega\hertz}, enabling us to achieve the highest possible loop gain with the chosen device.

To measure the achieved open-loop gain, we implemented a direct measurement feature within our phasemeter. 
A Gaussian noise generator with externally adjustable amplitude is included in the PL. 
When the associated switch is closed, Gaussian noise gets added to the servo output along with the initial frequency value.
Using an integrated logic aalyzer, we directly monitor the two signals before and after the Gaussian noise addition at the full processing speed.

\begin{figure}[!t]
\centering
\includegraphics[width=3.4in]{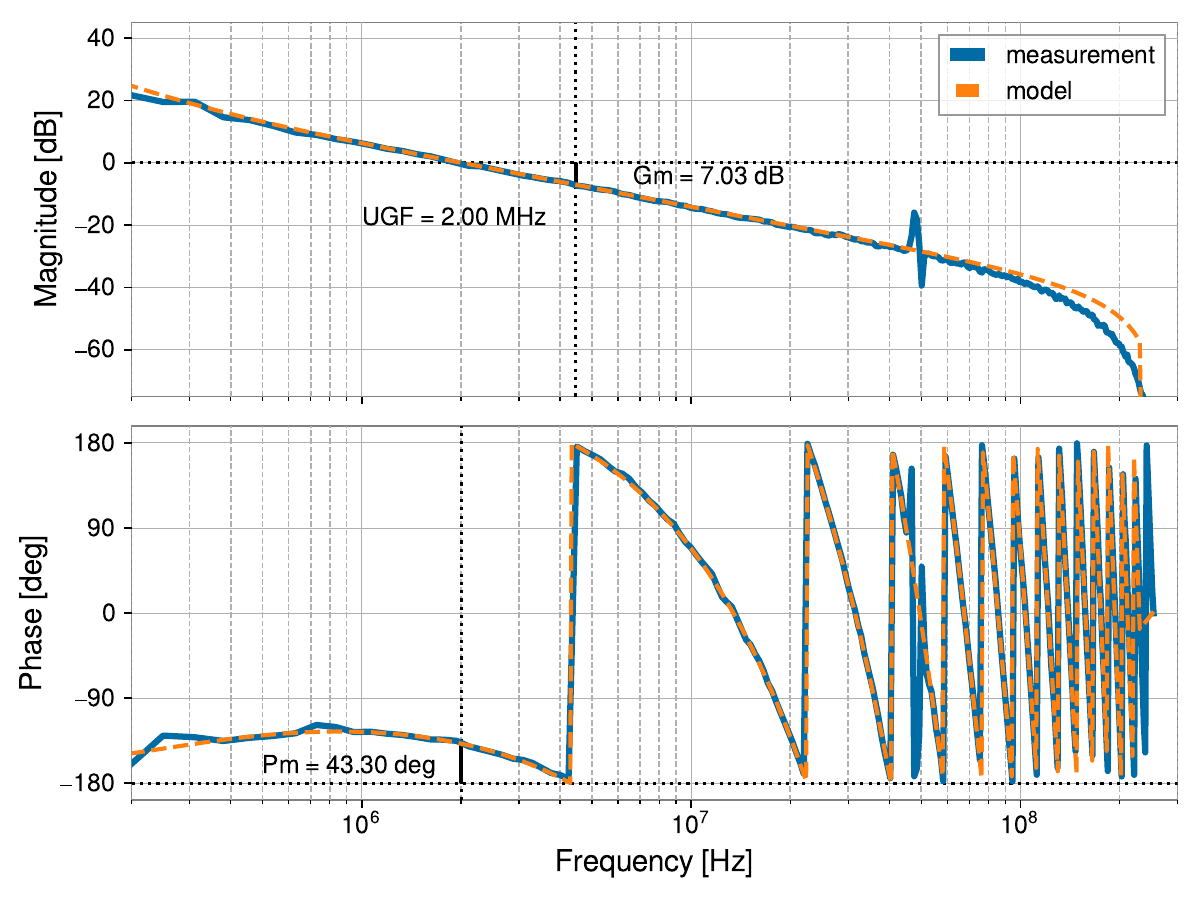}
\caption{Modeled dashed curve) and measured (solid curve) open-loop gain of the GHz Phasemeter. Both curves are in good agreement with each other. 
The peak appearing at \SI{48}{\mega\hertz} is the aliased second harmonic of the input signal. The deviation at the high-frequency end is not understood, but irrelevant for the analysis. 
We used `Spicypy'~\cite{spicypy} to get the measurement curve and `python-control'~\cite{python-control2021} for the loop modeling. UGF: unity gain frequency, Gm: gain margin, Pm: phase margin.}
\label{fig3}
\end{figure}

Using these two signals, it is then straightforward to compute the transfer function of the PLL. We use the Python package `Spicypy'~\cite{spicypy} for this purpose.
Fig.~\ref{fig3} shows the measured PLL open-loop gain against the modeled one. 
The PLL was configured to operate close to its highest stable loop bandwidth of \SI{2}{\mega\hertz}. 
The additional integral controller increases the loop gain even more rapidly below a corner frequency of \SI{300}{\kilo\hertz}, realizing a loop suppression of about 12 orders of magnitude at \SI{1}{\hertz}, which could be increased further with an additional integrator corner. 
One can easily tune this loop bandwidth to lower values by adapting the servo gains accordingly.
This new feature helps in tuning the tracking bandwidth for the realization of optimal tracking conditions based on the input signal dynamics.

\subsection{Residual Phase Error} \noindent %

\begin{figure*}[!t]
\centering
\subfloat[]{\includegraphics[width=3.4in]{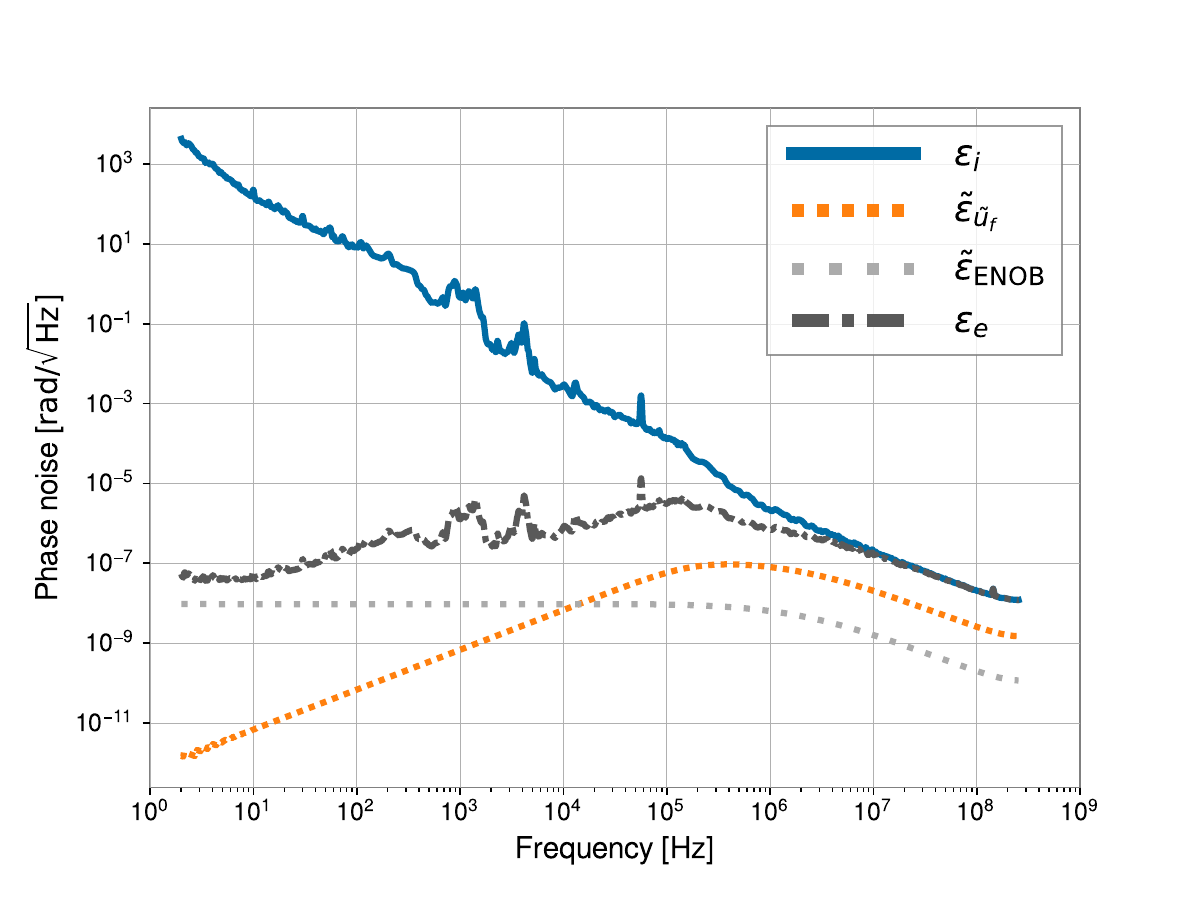}
\label{fig4a}}
\hfil
\subfloat[]{\includegraphics[width=3.4in]{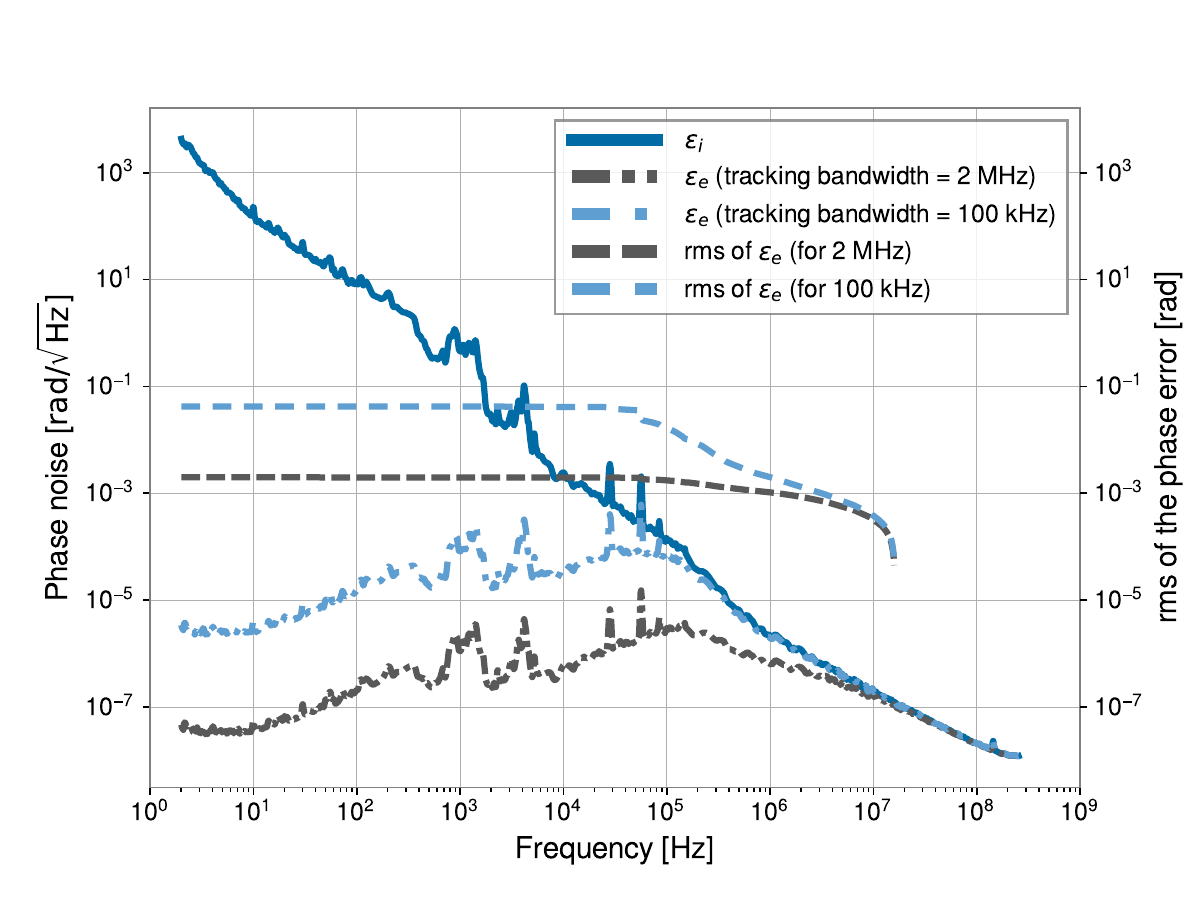}
\label{fig4b}}
\caption{(a) Measured phase noise spectrum of the beat note between two continuously tunable lasers from Toptica~\cite{toptica} solid curve) and its attenuation by a \SI{2}{\mega\hertz} loop bandwidth PLL (dash-dotted curve).
The spectrum is constructed using four measurements carried out in different Fourier frequency domains. 
The dotted curves represent effective phase noise contribution from the additional noise sources considered.
Notations used in the plot are according to the notations used in Fig.~\ref{fig2}. 
(b) Similar measurements carried out with two PLLs of different loop bandwidths. 
The loop attenuation of the input is calculated using the loop transfer function model. 
The dashed urves show the rms of this residual phase error, calculated starting from a Fourier frequency of \SI{16}{\mega\hertz} and going towards lower frequencies. 
}
\label{fig4}
\end{figure*}

The residual phase error within the ADPLL determines its stability and linearity. To calculate it, one has to consider the relevant spectra of each noise and the signal of interest and understand how they are suppressed by the loop transfer function. Then, one can calculate the rms value by integrating the spectra to get a simple metric for the loop. 
Fig.~\ref{fig4a} shows how the relevant noise spectra, shaped by the loop, contribute effectively to the residual phase noise error $\varepsilon_e$ of the PLL.

An $N$-bit signal, sampled with a sampling frequency of $f_s$, will result in truncation or quantization phase noise of~\cite{abramovici}
\begin{equation}
    \Tilde{u}_\mathrm{N} = \frac{2^{-N}}{\sqrt{6 \cdot f_s}} \hspace{0.2cm} \left[\frac{\SI{}{1}}{\sqrt{\SI{}{Hz}}}\right]\,.
    \label{eq:quant}
\end{equation}
The data converter has an effective number of bits above $10.5$, which takes account of the ADC noise floor as well~\cite{Liu2020}. 
This results in an effective phase noise contribution from this additive noise of 
\begin{equation}
    \Tilde{\varepsilon}_\mathrm{ENOB} = \frac{\Tilde{u}_\mathrm{10.5}}{A} \, \sqrt{2} \times \frac{G}{1+G} \hspace{0.2cm} \left[\frac{\SI{}{rad}}{\sqrt{\SI{}{Hz}}}\right]
\end{equation}
with $A$ as signal amplitude normalized to the maximum peak amplitude possible for these data converters and $G$ as the modeled open-loop gain of the PLL.

We also consider contributions from two truncations within the ADPLL that we introduced to reduce the overall bit-width, which also keeps the processing delays in check. 
One is the truncation of the $Q$ value, and the other is the truncation of the PIR value (see Fig.~\ref{fig2}).
Before such truncations, we also introduce dither to avoid nonwhite noise quantization errors, increasing the quantization noise in \ref{eq:quant}) by $(3)^{1/2}$.
We keep enough bits for the $Q$ truncation to ensure that its contribution is negligible compared to the ADC noise. For the PIR truncation, which couples with a different transfer function, we use 16 bits, which
keeps the residual phase error contribution $\Tilde{\varepsilon}_{{\Tilde{u}_f}}$ below \SI{1}{\micro\radian/(\hertz)^{1/2}} at all frequencies (see Fig.~\ref{fig4a}). 

To calculate the contribution from the signal, we use a spectrum of its dynamics that is suppressed by $1/(1+G)$. Here, we use the measured phase noise spectrum of a beat between two tunable lasers~\cite{toptica} as an example and plot both the spectrum with and without the loop suppression. It is evident that, for such conditions, the phase dynamics contribution dominates; hence, the goal is to maximize the tracking bandwidth (this could be different if, for example, the input signal has much higher additive noise~\cite{Sambridge2023}).
The resulting spectra and rms values of the residual phase error are shown in Fig.~\ref{fig4b} for two different bandwidths of the ADPLL, \SI{2}{\mega\hertz}, and \SI{100}{\kilo\hertz}.

Since lower rms values lead to a more stable and linear loop, we implemented a real-time measurement of the residual phase error.
For this, the fast $Q$ values first get pre-filtered to an intermediate frequency of \SI{32}{\mega\hertz} to suppress the second harmonic and other tones, and then $Q^2$ is computed inside the PL. 
It is then averaged using a first-order CIC filter and read out together with the other ADPLL values.  

\begin{figure}[!t]
\centering
\includegraphics[width=3.4in]{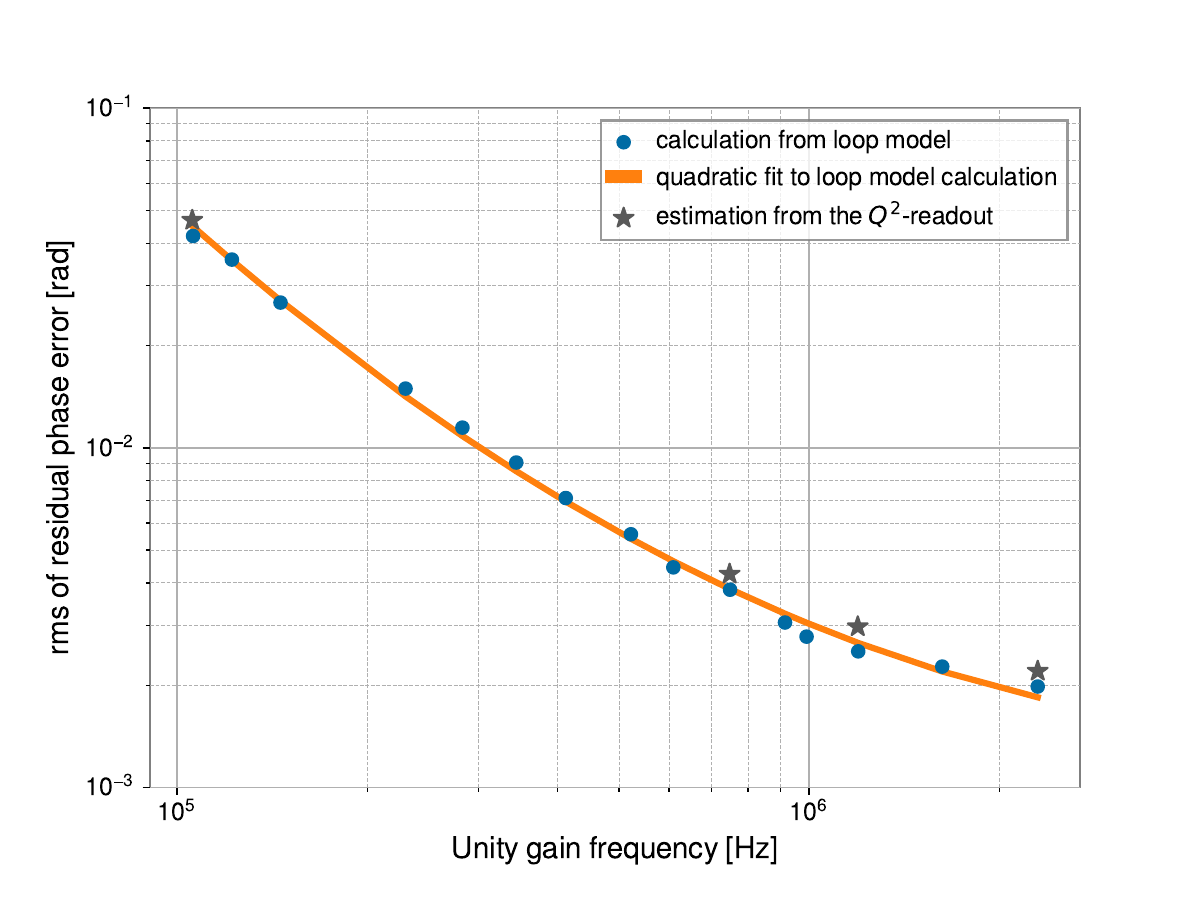}
\caption{rms of the residual phase error for different unity gain frequencies of the PLL. For a set of servo gain combinations, we calculated how the loop suppresses the laser phase noise spectrum shown in Fig.~\ref{fig4b}. rms of this residual phase error is calculated similarly and is marked as a dot in the graph. A quadratic fit in logarithmic scale to all those calculated points is shown as well. For four of the unity gain frequency conditions, we measured the residual phase error using the $Q^2$-readout.}
\label{fig7}
\end{figure}

By properly scaling these $Q^2$ values into units of radians, this readout can be used to estimate the residual phase error of the loop directly. 
The results are shown in Fig.~\ref{fig7}, where we can see a strong agreement between the modeled residual phase error and the one predicted using $Q^2$-readout.
This feature enables us to deduce and minimize the residual phase error while tracking a signal by tuning the PLL servo gains, which is especially useful when a model of the signal dynamics is not available or if the spectrum of the dynamics is non-stationary. One should note that the lasers used for our experiments have a rather non-stationary noise because they are sensitive to acoustics and vibrations. We isolated the laser from acoustics and vibrations for the measurements and used a quiet stretch of the measurement time to estimate the spectrum. Hence, the shown spectra should be considered as a minimal noise level of these lasers, which could also explain differences between the predicted and the modeled residual phase error in Fig.~\ref{fig7}.

\section{Performance}
\label{sec:performance} \noindent %
After conducting functional tests of our phasemeter, with a focus on testing the ADPLL using behavioral simulations, we implemented the algorithms into the ZCU111 evaluation board and conducted performance measurements using analog signals. 

\subsection{Measurement Noise Floor} 
\begin{figure*}[!t]
\centering
\subfloat[]{\includegraphics[width=3.4in]{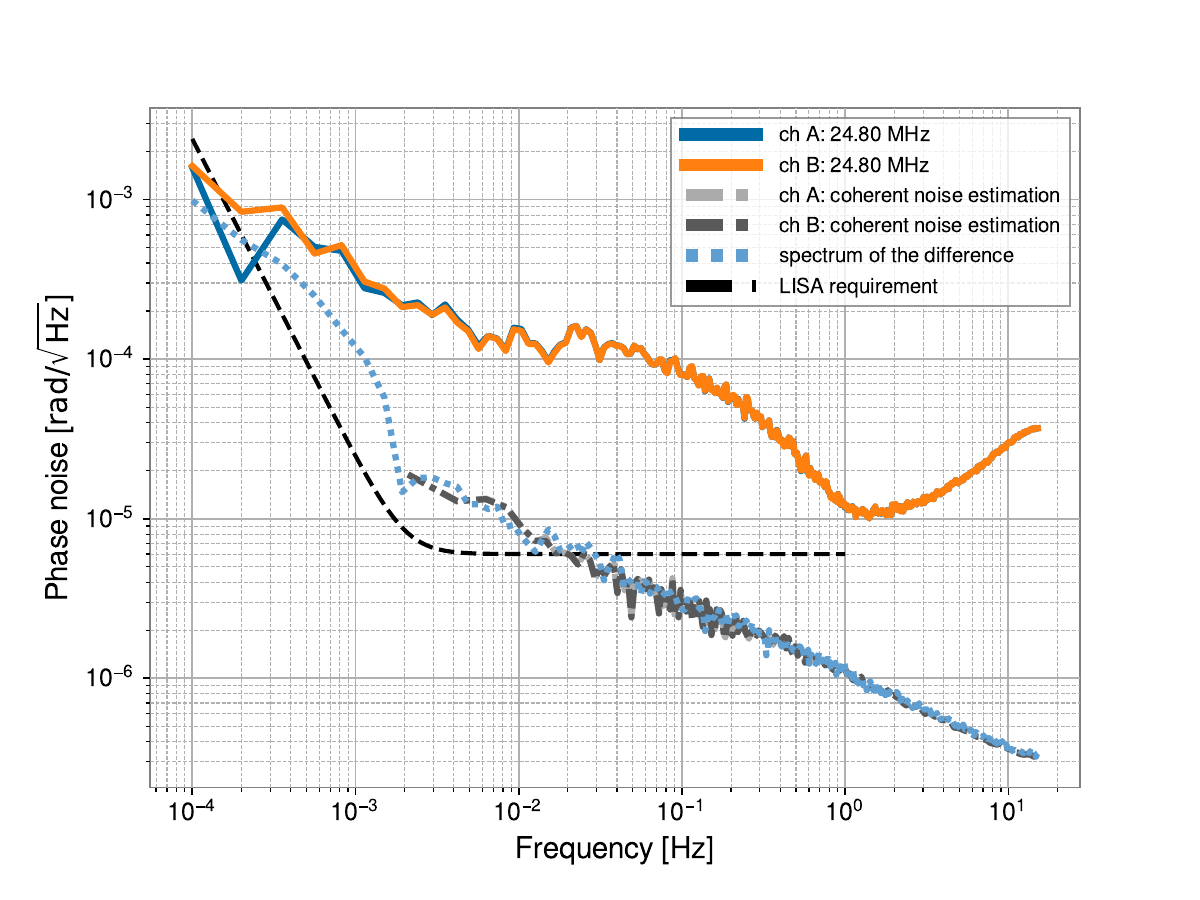}
\label{fig5a}}
\hfil
\subfloat[]{\includegraphics[width=3.4in]{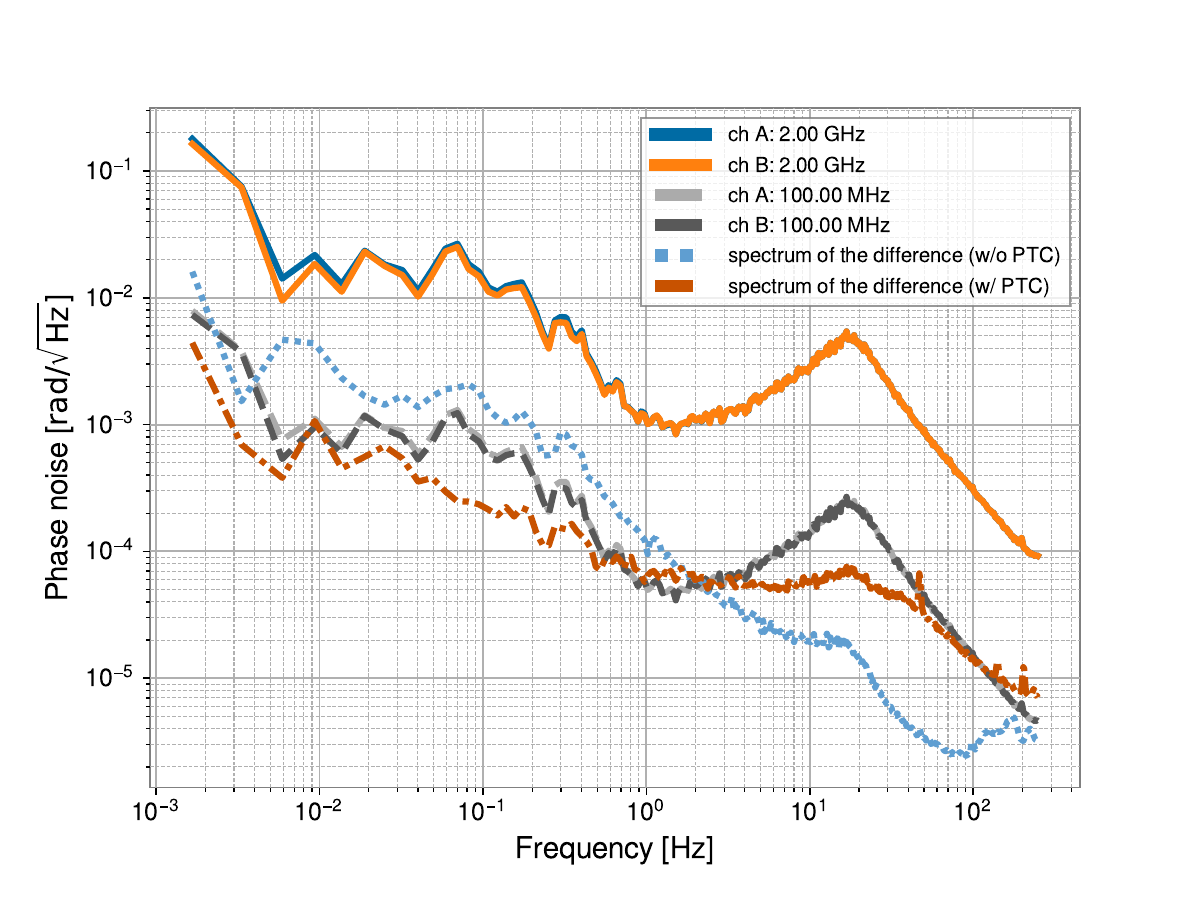}
\label{fig5b}}
\caption{(a) Self-noise levels of two phasemeter channels calculated using coherent subtraction method with transfer function estimation (dash-dotted curves). The input signal contained a \SI{24.8}{\mega\hertz} main tone.
Both channels have a very similar measurement noise floor. 
In the zero-measurement technique, we subtract the two time series and evaluate the spectrum of the difference (dotted curve).
Above a Fourier frequency of about \SI{0.02}{\hertz}, the phase noise floor is below the LISA requirement (dashed curve). 
(b) Two PLLs were set to track a \SI{2}{\giga\hertz} main tone and a \SI{100}{\mega\hertz} pilot tone. While post-processing, each channel's main tone phase readout values were corrected using the corresponding pilot phase. Above a Fourier frequency of about \SI{0.1}{\hertz}, the phase noise floor is in the sub-milli radian regime. PTC: pilot tone correction.}
\label{fig5}
\end{figure*}

\noindent %
To probe the phase measurement noise floor, two tones of different frequencies were generated using an external signal generator.
These tones are initially combined and then split using RF power splitters/combiners before they are fed into two phasemeter channels, A and B. Each channel uses one ADPLL to track the signal and the other to track the pilot tone. The PIR values of each ADPLL were then used to calculate the measured phase for each channel. Comparing two channels, one can determine the phase measurement noise floor, so-called zero-measurement. 
Typically, this is done by simply subtracting the phase values from the two channels and then computing the spectral density, providing a measure of the incoherent sum of the noise in both channels. Here, we also estimate the noise of each channel by using a coherent subtraction method with transfer function estimation~\cite{Kirchhoff2017}. We compare the results from both methods for the tracking of a tone in the LISA band in Fig.~\ref{fig5a}. Both methods estimate similar noise levels, indicating no significant phase noise differences between the channels. Phase noise of the analog front-end, often driven by temperature and humidity fluctuations~\cite{Liu2014}, is one possible candidate for the current limitations but was not studied further. 

It is important here to note that the input signal is split and fed into two channels just for the estimation of the measurement noise floor.
For the general functioning of the phasemeter, e.g., when tracking an RF signal, one single channel of the phasemeter is sufficient.
Thus, the eight-channel GHz Phasemeter gives, in general, the flexibility of tracking eight independent input signals.

ADC sampling jitter is another prominent limitation in digital phasemeters that leads to a frequency-dependent phase noise, which can be reduced with a pilot tone correction~\cite{og-thesis}. To demonstrate this, we used the second frequency tracked in each channel to calculate pilot-tone corrected phase values and then again calculate the spectral density of the corrected difference. 
For LISA band measurements in Fig.~\ref{fig5a}, we found no performance improvement when using an additional \SI{38.2}{\mega\hertz} pilot tone and the corresponding correction (not shown). Instead, the phase noise keeps its $1/(f)^{1/2}$ shape, indicating that an unknown Flicker-type noise is the current limit for a LISA-type phasemeter implementation in these RFSoCs.
The results in Fig.~\ref{fig5b} indicate that currently, we are limited by ADC timing jitter at frequencies below \SI{2}{\hertz} for higher signal frequencies. For frequencies up to \SI{2}{\giga\hertz}, we reach a noise floor below \SI{1}{\milli\radian/(\hertz)^{1/2}} at almost all readout frequencies without optimizing the analog front-end.

\subsection{Role of the High Tracking Bandwidth} \noindent %
To demonstrate the advantage of having high loop (or tracking) bandwidth, we simultaneously track the beat note between two free-running external cavity diode lasers under two different tracking bandwidth conditions. As mentioned previously, these lasers have a rather high phase noise spectrum and are prone to non-stationary excess noise, which makes proper phase-tracking difficult using commercially available phasemeter solutions or a LISA-type phasemeter. 
We tracked the laser beat simultaneously using two ADPLLs operating on one ADC channel with different servo gains, creating the two tracking bandwidth conditions also studied in Fig.~\ref{fig4b}. 

\begin{figure}[!t]
\centering
\includegraphics[width=3.4in]{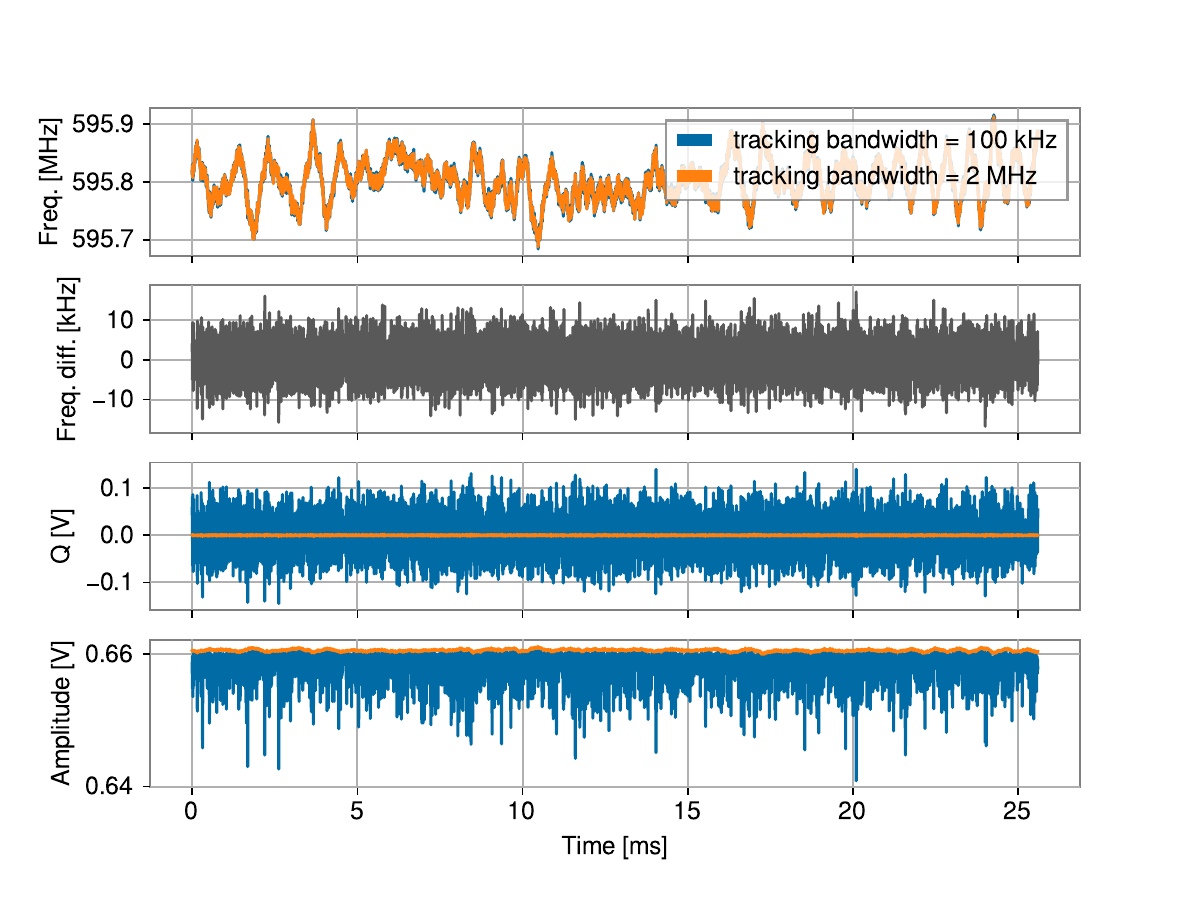}
\caption{Beat note between the two lasers was tracked with the GHz Phasemeter, once with a tracking bandwidth of \SI{100}{\kilo\hertz} and the other time with \SI{2}{\mega\hertz}. }
\label{fig6}
\end{figure}

The measured frequency, $Q$, and amplitude time series are shown in Fig.~\ref{fig6}, together with the difference between the two frequency estimates from high and low loop gain measurements. 
Looking at their behavior, the role of high tracking bandwidth becomes evident.
The tracking with low loop bandwidth introduces significant frequency errors, and for $Q$, this results in the expected higher rms. For the amplitude, one finds an underestimation and one-sided noise behavior that results from a non-linear effect in the $I$ demodulation and decimation~\cite{og-thesis}.
Limited tracking bandwidth can also cause the ADPLL to lose track of the input signal for a moment or permanently. 

With the \SI{2}{\mega\hertz} tracking bandwidth, we can measure the beat note between these non-stabilized lasers without phase errors, cycle slips, or loss of lock for many hours, with no indication that longer locks are not possible.

\subsection{Tracking Highly Dynamic Signals} \noindent %
A PLL can track signals only up to certain dynamics. The maximum tracking speed of a PLL defines the maximum rate of change of input frequency that the PLL can track reliably. To characterize this value for the GHz Phasemeter, initially, it was locked to a fixed frequency of an external signal generator. 
Then, by introducing a frequency modulation at different modulation frequencies, we tested the behavior of the phasemeter. 
With a modulation frequency of \SI{30}{\kilo\hertz} and a frequency deviation of \SI{4}{\mega\hertz}, the GHz Phasemeter could track the frequency changes reliably, showing that it can track signals that change at a speed of \SI{240}{\giga\hertz/\second}. 
It is worth noting that this is not the maximum tracking speed of the ADPLL but that it was limited by the available signal generators and the data readout frequency of the phasemeter.  

For locking the phasemeter to a highly dynamic signal, the acquisition range, also known as the pull-in range or capture range of a PLL is critical.
It defines how far the initial frequency $f_0$ can deviate from the input signal frequency for a successful lock of the PLL.
Using a straightforward testing method, where $f_0$ is kept at a fixed value, and by deviating the input tone from this value, we found that our PLL has an acquisition range of \SI{4.1}{\mega\hertz}. Since the acquisition range scales with the tracking bandwidth, it is beneficial that our phasemeter provides both high tracking and detection bandwidth. This should enable us to use a Fast Fourier Transform of the ADC data to identify the desired locking frequency in an automated lock acquisition procedure as developed for space-based laser interferometry~\cite{Bachman2017}.

\section{Conclusion} \noindent %
In this work, we have developed and demonstrated an eight-channel GHz Phasemeter based on an RFSoC system. 
The GHz Phasemeter has a signal bandwidth of \SI{2.048}{\giga\hertz}, a tracking bandwidth of \SI{2}{\mega\hertz}, a capture range of \SI{4.1}{\mega\hertz}, and a tracking speed of more than \SI{240}{\giga\hertz/s}, setting a new benchmark for these parameters.
These make it, to the best of our knowledge, the highest bandwidth, highly stable, and the fastest phase-tracking instrument reported so far. 
The two additional novel features of our phasemeter, direct measurement of tracking bandwidth and direct estimation of residual phase error, help in realizing optimal tracking conditions based on the input signal dynamics.
With this phasemeter, we can track the beat note between noisy lasers, and we can use it to realize laser interferometric readout schemes that require a very high range and dynamics of frequency change.

In the future, a calibration of the phase response and transfer function of our device will be investigated to enable absolute phase measurements. Our approach can be extended with techniques like ADC frequency interleaving to push the phasemeter detection bandwidth to even higher levels. Then, it will be possible to use ultra-high dynamic range frequency measurements of the digital phasemeters in even more applications.

\section*{Acknowledgment}
\noindent The authors thank the Deutsches Elektronen-Synchrotron (DESY) Maschine Strahlkontrollen (MSK) group for their FPGA infrastructure, which was made use of in this project.
They also thank David Shoemaker for his comments on the draft of this manuscript and Jonathan Carter for his suggestions during the LIGO P\&P review process.

\bibliography{references}

\begin{thebibliography}{10}
\providecommand{\url}[1]{#1}
\csname url@samestyle\endcsname
\providecommand{\newblock}{\relax}
\providecommand{\bibinfo}[2]{#2}
\providecommand{\BIBentrySTDinterwordspacing}{\spaceskip=0pt\relax}
\providecommand{\BIBentryALTinterwordstretchfactor}{4}
\providecommand{\BIBentryALTinterwordspacing}{\spaceskip=\fontdimen2\font plus
\BIBentryALTinterwordstretchfactor\fontdimen3\font minus
  \fontdimen4\font\relax}
\providecommand{\BIBforeignlanguage}[2]{{%
\expandafter\ifx\csname l@#1\endcsname\relax
\typeout{** WARNING: IEEEtran.bst: No hyphenation pattern has been}%
\typeout{** loaded for the language `#1'. Using the pattern for}%
\typeout{** the default language instead.}%
\else
\language=\csname l@#1\endcsname
\fi
#2}}
\providecommand{\BIBdecl}{\relax}
\BIBdecl

\bibitem{Pollack2006}
\BIBentryALTinterwordspacing
S.~E. Pollack and R.~T. Stebbins, ``Demonstration of the zero-crossing
  phasemeter with a lisa test-bed interferometer,'' \emph{Classical and Quantum
  Gravity}, vol.~23, no.~12, p. 4189–4200, May 2006. [Online]. Available:
  \url{http://dx.doi.org/10.1088/0264-9381/23/12/014}
\BIBentrySTDinterwordspacing

\bibitem{6230960}
S.~Johansson, ``New frequency counting principle improves resolution,'' in
  \emph{Proceedings of the 20th European Frequency and Time Forum}, 2006, pp.
  139--146.

\bibitem{Shaddock2006}
\BIBentryALTinterwordspacing
D.~Shaddock, B.~Ware, P.~G. Halverson, R.~E. Spero, and B.~Klipstein,
  ``Overview of the lisa phasemeter,'' in \emph{AIP Conference
  Proceedings}.\hskip 1em plus 0.5em minus 0.4em\relax AIP, 2006. [Online].
  Available: \url{http://dx.doi.org/10.1063/1.2405113}
\BIBentrySTDinterwordspacing

\bibitem{Gerberding2013}
\BIBentryALTinterwordspacing
O.~Gerberding, B.~Sheard, I.~Bykov, J.~Kullmann, J.~J.~E. Delgado, K.~Danzmann,
  and G.~Heinzel, ``Phasemeter core for intersatellite laser heterodyne
  interferometry: modelling, simulations and experiments,'' \emph{Classical and
  Quantum Gravity}, vol.~30, no.~23, p. 235029, Nov. 2013. [Online]. Available:
  \url{https://doi.org/10.1088/0264-9381/30/23/235029}
\BIBentrySTDinterwordspacing

\bibitem{moku}
\url{https://www.liquidinstruments.com/products/integrated-instruments/phasemeter/},
  [Online; accessed 22-Jan-2024].

\bibitem{Tudosa2022}
\BIBentryALTinterwordspacing
I.~Tudosa, F.~Picariello, P.~Daponte, L.~De~Vito, S.~Rapuano, and N.~G.
  Paulter, ``Prototype of high accuracy single input phase measurement
  instrument,'' \emph{Measurement}, vol. 201, p. 111595, Sep. 2022. [Online].
  Available: \url{http://dx.doi.org/10.1016/j.measurement.2022.111595}
\BIBentrySTDinterwordspacing

\bibitem{Chaudhary2015}
\BIBentryALTinterwordspacing
S.~Chaudhary and A.~Samant, ``Characterization and calibration techniques for
  multi-channel phase-coherent systems,'' in \emph{2015 IEEE
  AUTOTESTCON}.\hskip 1em plus 0.5em minus 0.4em\relax IEEE, Nov. 2015, p.
  334–338. [Online]. Available:
  \url{http://dx.doi.org/10.1109/AUTEST.2015.7356512}
\BIBentrySTDinterwordspacing

\bibitem{Schwarze2019}
\BIBentryALTinterwordspacing
T.~S. Schwarze, G.~Fernández~Barranco, D.~Penkert, M.~Kaufer, O.~Gerberding,
  and G.~Heinzel, ``Picometer-stable hexagonal optical bench to verify lisa
  phase extraction linearity and precision,'' \emph{Physical Review Letters},
  vol. 122, no.~8, Feb. 2019. [Online]. Available:
  \url{http://dx.doi.org/10.1103/PhysRevLett.122.081104}
\BIBentrySTDinterwordspacing

\bibitem{1095423}
G.~Ascheid and H.~Meyr, ``Cycle slips in phase-locked loops: A tutorial
  survey,'' \emph{IEEE Transactions on Communications}, vol.~30, no.~10, pp.
  2228--2241, 1982.

\bibitem{Hsu2010}
\BIBentryALTinterwordspacing
M.~T.~L. Hsu, I.~C.~M. Littler, D.~A. Shaddock, J.~Herrmann, R.~B. Warrington,
  and M.~B. Gray, ``Subpicometer length measurement using heterodyne laser
  interferometry and all-digital rf phase meters,'' \emph{Optics Letters},
  vol.~35, no.~24, p. 4202, Dec. 2010. [Online]. Available:
  \url{http://dx.doi.org/10.1364/OL.35.004202}
\BIBentrySTDinterwordspacing

\bibitem{Sambridge2023}
\BIBentryALTinterwordspacing
C.~S. Sambridge, L.~E. Roberts, A.~R. Wade, J.~T. Valliyakalayil, E.~R. Rees,
  N.~Chabbra, J.~Zhang, A.~J. Sutton, D.~A. Shaddock, and K.~McKenzie,
  ``Subfemtowatt laser phase tracking,'' \emph{Physical Review Letters}, vol.
  131, no.~19, Nov. 2023. [Online]. Available:
  \url{http://dx.doi.org/10.1103/PhysRevLett.131.193804}
\BIBentrySTDinterwordspacing

\bibitem{vanHeijningen2023}
\BIBentryALTinterwordspacing
J.~V. van Heijningen, H.~J.~M. ter Brake, O.~Gerberding, S.~C. Subrahmanya,
  J.~Harms, X.~Bian, A.~Gatti, M.~Zeoli, A.~Bertolini, C.~Collette, A.~Perali,
  N.~Pinto, M.~Sharma, F.~Tavernier, and J.~Rezvani, ``The payload of the lunar
  gravitational-wave antenna,'' \emph{Journal of Applied Physics}, vol. 133,
  no.~24, Jun. 2023. [Online]. Available:
  \url{https://doi.org/10.1063/5.0144687}
\BIBentrySTDinterwordspacing

\bibitem{Eichholz2015}
\BIBentryALTinterwordspacing
J.~Eichholz, D.~B. Tanner, and G.~Mueller, ``Heterodyne laser frequency
  stabilization for long baseline optical interferometry in space-based
  gravitational wave detectors,'' \emph{Physical Review D}, vol.~92, no.~2,
  Jul. 2015. [Online]. Available:
  \url{http://dx.doi.org/10.1103/PhysRevD.92.022004}
\BIBentrySTDinterwordspacing

\bibitem{Carter2020}
\BIBentryALTinterwordspacing
J.~Carter, S.~Kohlenbeck, P.~Birckigt, R.~Eberhardt, G.~Heinzel, and
  O.~Gerberding, ``A high q, quasi-monolithic optomechanical inertial sensor,''
  in \emph{2020 IEEE International Symposium on Inertial Sensors and Systems
  (INERTIAL)}.\hskip 1em plus 0.5em minus 0.4em\relax IEEE, Mar. 2020.
  [Online]. Available:
  \url{http://dx.doi.org/10.1109/INERTIAL48129.2020.9090085}
\BIBentrySTDinterwordspacing

\bibitem{Hines2023}
\BIBentryALTinterwordspacing
A.~Hines, A.~Nelson, Y.~Zhang, G.~Valdes, J.~Sanjuan, and F.~Guzman, ``Compact
  optomechanical accelerometers for use in gravitational wave detectors,''
  \emph{Applied Physics Letters}, vol. 122, no.~9, Feb. 2023. [Online].
  Available: \url{http://dx.doi.org/10.1063/5.0142108}
\BIBentrySTDinterwordspacing

\bibitem{Liu2021}
\BIBentryALTinterwordspacing
H.-S. Liu, T.~Yu, and Z.-R. Luo, ``A low-noise analog frontend design for the
  taiji phasemeter prototype,'' \emph{Review of Scientific Instruments},
  vol.~92, no.~5, May 2021. [Online]. Available:
  \url{http://dx.doi.org/10.1063/5.0042249}
\BIBentrySTDinterwordspacing

\bibitem{Liu2014}
\BIBentryALTinterwordspacing
H.-S. Liu, Y.-H. Dong, Y.-Q. Li, Z.-R. Luo, and G.~Jin, ``The evaluation of
  phasemeter prototype performance for the space gravitational waves
  detection,'' \emph{Review of Scientific Instruments}, vol.~85, no.~2, Feb.
  2014. [Online]. Available: \url{http://dx.doi.org/10.1063/1.4865121}
\BIBentrySTDinterwordspacing

\bibitem{Isleif2014}
\BIBentryALTinterwordspacing
K.-S. Isleif, O.~Gerberding, S.~K\"{o}hlenbeck, A.~Sutton, B.~Sheard,
  S.~Goßler, D.~Shaddock, G.~Heinzel, and K.~Danzmann, ``Highspeed multiplexed
  heterodyne interferometry,'' \emph{Optics Express}, vol.~22, no.~20, p.
  24689, Oct. 2014. [Online]. Available:
  \url{http://dx.doi.org/10.1364/OE.22.024689}
\BIBentrySTDinterwordspacing

\bibitem{rfsoc}
\url{https://www.xilinx.com/products/boards-and-kits/zcu111.html}, [Online;
  accessed 25-Jan-2024].

\bibitem{spicypy}
\BIBentryALTinterwordspacing
A.~Basalaev, C.~Darsow-Fromm, O.~Gerberding, M.~Hewitson, A.~Patra,
  C.~Mow-Lowry, D.~Voigt, N.~Holland, N.~L. G\"{o}bbels, O.~Vega, P.~Saffarieh,
  S.~Bania, and S.~Chalathadka~Subrahmanya, ``Spicypy,'' 2023. [Online].
  Available: \url{https://zenodo.org/doi/10.5281/zenodo.10033638}
\BIBentrySTDinterwordspacing

\bibitem{python-control2021}
\BIBentryALTinterwordspacing
S.~Fuller, B.~Greiner, J.~Moore, R.~Murray, R.~van Paassen, and R.~Yorke, ``The
  python control systems library (python-control),'' in \emph{2021 60th IEEE
  Conference on Decision and Control (CDC)}.\hskip 1em plus 0.5em minus
  0.4em\relax IEEE, Dec. 2021. [Online]. Available:
  \url{http://dx.doi.org/10.1109/CDC45484.2021.9683368}
\BIBentrySTDinterwordspacing

\bibitem{abramovici}
A.~Abramovici and J.~Chapsky, \emph{Feedback Control Systems: A Fast-Track
  Guide for Scientists and Engineers}.\hskip 1em plus 0.5em minus 0.4em\relax
  Springer Science+Business Media, LLC, 2000.

\bibitem{Liu2020}
\BIBentryALTinterwordspacing
C.~Liu, M.~E. Jones, and A.~C. Taylor, ``Characterizing the performance of
  high-speed data converters for rfsoc-based radio astronomy receivers,''
  \emph{Monthly Notices of the Royal Astronomical Society}, vol. 501, no.~4, p.
  5096–5104, Dec. 2020. [Online]. Available:
  \url{http://dx.doi.org/10.1093/mnras/staa3895}
\BIBentrySTDinterwordspacing

\bibitem{toptica}
\url{https://www.toptica.com/products/tunable-diode-lasers/ecdl-dfb-lasers/ctl},
  [Online; accessed 12-Feb-2024].

\bibitem{Kirchhoff2017}
\BIBentryALTinterwordspacing
R.~Kirchhoff, C.~M. Mow-Lowry, V.~B. Adya, G.~Bergmann, S.~Cooper, M.~M. Hanke,
  P.~Koch, S.~M. K\"{o}hlenbeck, J.~Lehmann, P.~Oppermann, J.~W\"{o}hler, D.~S.
  Wu, H.~L\"{u}ck, and K.~A. Strain, ``Huddle test measurement of a near
  johnson noise limited geophone,'' \emph{Review of Scientific Instruments},
  vol.~88, no.~11, Nov. 2017. [Online]. Available:
  \url{http://dx.doi.org/10.1063/1.5000592}
\BIBentrySTDinterwordspacing

\bibitem{og-thesis}
O.~Gerberding, ``Phase readout for satellite interferometry,'' Ph.D.
  dissertation, Gottfried Wilhelm Leibniz Universit\"at Hannover, 2014.

\bibitem{Bachman2017}
\BIBentryALTinterwordspacing
B.~Bachman, G.~de~Vine, J.~Dickson, S.~Dubovitsky, J.~Liu, W.~Klipstein,
  K.~McKenzie, R.~Spero, A.~Sutton, B.~Ware, and C.~Woodruff, ``Flight
  phasemeter on the laser ranging interferometer on the grace follow-on
  mission,'' \emph{Journal of Physics: Conference Series}, vol. 840, p. 012011,
  May 2017. [Online]. Available:
  \url{http://dx.doi.org/10.1088/1742-6596/840/1/012011}
\BIBentrySTDinterwordspacing

\end{thebibliography}
\bibliographystyle{IEEEtran}

\begin{IEEEbiography}[{\includegraphics[width=1in,height=1.25in,clip,keepaspectratio]{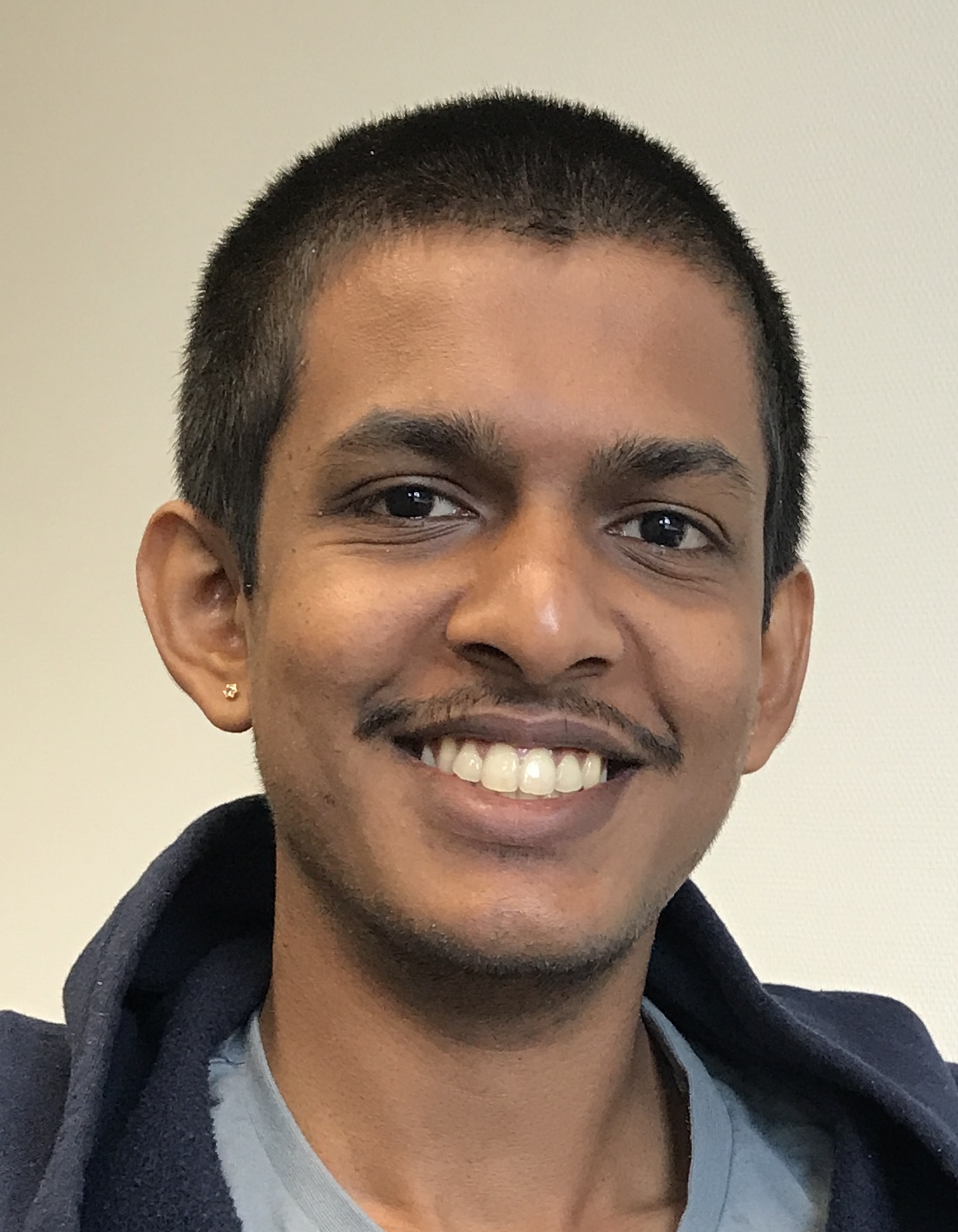}}]{Shreevathsa Chalathadka Subrahmanya}
received the B.Sc. and M.Sc. degrees in physics from Mangalore University, Mangaluru, Karnataka, India, in 2017 and 2019, respectively. 
He is currently pursuing the Ph.D. degree with the Institute of Experimental Physics, University of Hamburg, Hamburg, Germany.

He is working on interferometric-based high-precision displacement sensors and FPGA-based readout-system development for the same. 
His research interests include gravitational wave detection, metrology, FPGA systems, and interferometry.

\end{IEEEbiography}

\begin{IEEEbiography}[{\includegraphics[width=1in,height=1.25in,clip,keepaspectratio]{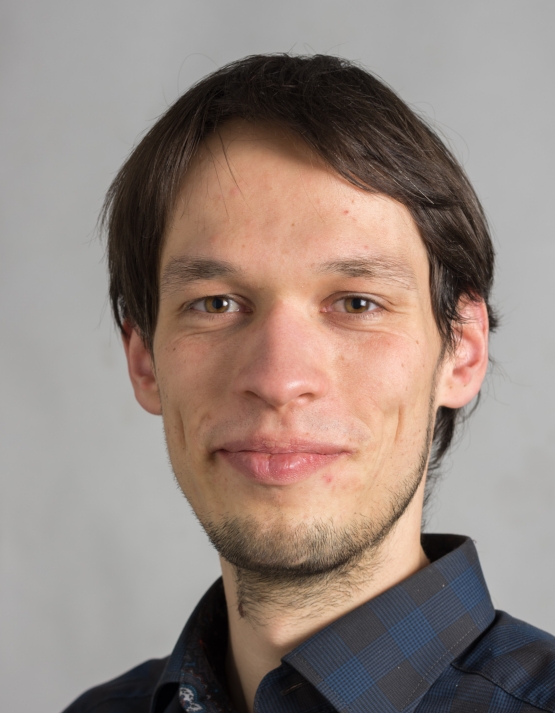}}]{Christian Darsow-Fromm}
received the B.Sc. degree in computing in science and the M.Sc. and Ph.D. degrees in physics from the University of Hamburg, Hamburg, Germany, in 2014, 2016, and 2022, respectively.

Currently, he is a Post-Doctoral Researcher at the Institute of Experimental Physics, University of Hamburg.
He works on the development of a ground-support equipment phasemeter for the space-based gravitational-wave detector, LISA.

\end{IEEEbiography}

\begin{IEEEbiography}[{\includegraphics[width=1in,height=1.25in,clip,keepaspectratio]{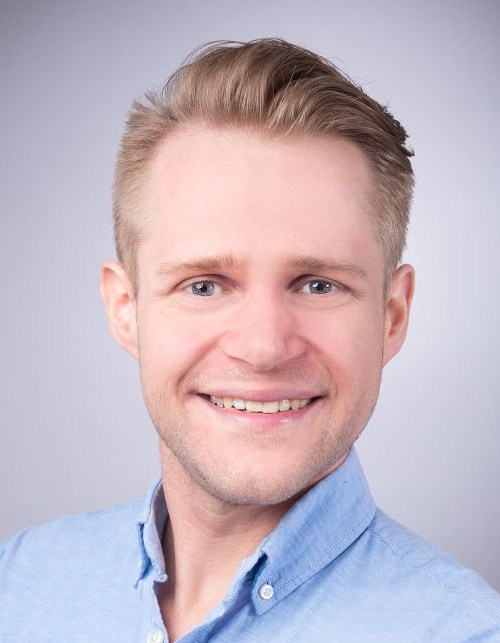}}]{Oliver Gerberding}
received the Diploma degree in technical physics and the Ph.D. degree in physics from the Leibniz Universtät Hannover, Hannover, Germany, in 2009 and 2014, respectively.

He currently leads a Research Group for Gravitational Wave Detection, Institute of Experimental Physics, University of Hamburg, Hamburg, Germany.
He studies technologies for ground and space-based gravitational-wave detectors, with a focus on laser interferometry, FPGA-based readout systems, compact sensors, scattered light and the reduction of low-frequency noise sources.

\end{IEEEbiography}

\end{document}